\documentclass{article}[11pt]
\usepackage[all]{xy}

\def\P{\mathcal P\/}
\def\S{\mathcal S\/}
\def\vth{\vartheta}
\def\vph{\varphi}
\def\A{\mathcal A\/}
\def\T{\mathcal T\/}
\def\J{J_1\/(\V)}
\def\V{\mathcal V\/}
\def\={\, = \,}
\def\dfrac{\displaystyle \frac}
\def\dx#1{\dot{x}\,_{#1}}
\def\dy#1{\dot{y}\,_{#1}}
\def\dotp#1{\dot{\psi}\,_{#1}}
\def\dott#1{\dot{\vth}\,_{#1}}
\def\dotf#1{\dot{\vph}\,_{#1}}

\begin{document}

 \vskip-2cm

\title{Some results on ideal impacts of billiard balls
         }
\author{Stefano Pasquero \\
        Dipartimento di Matematica dell'Universit\`a di Parma \\
        Viale G.P. Usberti, 53/A - 43100 Parma (Italia) \\
        E-mail: stefano.pasquero@unipr.it
        }

\maketitle

\date{}

\begin{abstract}
\noindent We analyze the impact of two equal billiard balls in three ideal situations: when the
balls freely slide on the plane of the billiard, when they roll without sliding and when one of
them freely slides and the other rolls. In all the cases we suppose that the contact between the
balls is smooth. We base our analysis on some recent general theoretical results on ideal impacts
obtained by means of Differential Geometric Impulsive Mechanics. We use symbolic computation
software to solve the computational difficulties arising by the high number of degrees of freedom
of the system. Some particular but significative impacts, with opportunely assigned left
velocities and positions of the balls, are analyzed in details. The results admit easy
interpretations that turn out to be in good agreement with the reasonable forecasts and the
behaviours of real systems.
\end{abstract}

\vskip 0.5 truecm
\par \par \noindent {\bf PACS:} 03.20+i
\newline
{\bf 2000 Mathematical subject classification:} 70F35, 70F99
\newline
{\bf Keywords:} billiard, impulsive reaction.

\thispagestyle{empty}

\section*{Introduction}

The impact problems arising in the study of billiard games has always represented, for their
possibility of being modelled with a lower or higher degree of simplification and the facility of
comparison of the theoretical results with the real behaviour of the balls in the game,
interesting challenges and severe tests for all who study Classical Impulsive Mechanics. The
geometric approach to these mechanical problems has recently given important theoretical results,
when techniques of modern Differential Geometry were fruitfully applied to investigate the concept
of ideality of unilateral constraints (\cite{Pasquero2005uni}), even in presence of kinetic
constraints (\cite{Pasquero2006}). Unfortunately, because of computational difficulties due to the
number of degrees of freedom of the systems, often theoretical results cannot be straightly
applied to analyze practical problems: the impact between two equal billiard balls can be modelled
with an impulsive mechanical system with 10 degrees of freedom (two for the position of the center
of mass and three for the Euler angles for each ball) simultaneously subject to an instantaneous
unilateral constraint (the contact condition), possibly together with permanent and/or
instantaneous kinetic constraints (the rolling conditions), all of them involving several
parameters. The calculations arising in the study of this impact are almost prohibitive if
approached with hand techniques. Luckily, nowadays the use of symbolic computation softwares can
give a very great help to solve (part of) these problems.

In this paper we present the application of the theoretical results of \cite{Pasquero2005uni,
Pasquero2006} to the classical example of impact of identical billiard balls in three ideal but
nevertheless significative cases: the smooth case, when both the balls slide on the horizontal
plane; the rolling case, when both the balls roll without sliding on the horizontal plane; a mixed
case, when a rolling ball impacts with a ball that slides. In everyone of these cases, the theory
is powerful enough to allow a very general approach, with both the balls moving on the plane, and
with general orientations of the balls at the impact. Even in this general situation, some results
(for example concerning the angle formed by the directions of the balls after the impact) will
appear in a very clear form. However, in order to have a clearer interpretation of some results,
it will be almost unavoidable the analysis of particular impacts. This choice will appear even
more necessary in the mixed case, if we want to avoid almost meaningless situations. In these
cases, the results assume very simple aspects, that can be easily compared with those of well
known experiences.

In the first and simplest case, we had to solve a $11\times 11$ linear system (fortunately with a
very simple structure), while in the second and more complicated case we had to solve a  $15\times
15$ linear system, both involving several parameters, such as the trigonometric functions of the
Euler angles of the balls: to this aim, we used CoCoA 4.7, a freeware software for exact symbolic
computation (\cite{CoCoA}).

We choose to focus our attention to the most meaningful results from the point of view of the
physical behaviour of the balls. Then the computational aspect of the analysis is not treated in
details in the paper. Moreover, we point out that, once the results of
\cite{Pasquero2005uni,Pasquero2006} are assumed, the algorithm to determine the behaviour of the
balls after the impact starting by the knowledge of positions and motions of the balls before the
impact (is based on but) do not involve refined geometric techniques. Then, for the same reason
written above, we describe in details the algorithm (that is, we describe in details the results
of \cite{Pasquero2005uni,Pasquero2006}), and then we analyze in details the results obtained, but
we expose only a very concise description of the geometry underlying the problem.

\section{Preliminaries}

The theoretical foundation of ideality for unilateral constraints acting on a mechanical system
with a finite number of degrees of freedom can be fruitfully framed in the context of Modern
Differential Geometry, in particular using the concepts of fibre bundles, their jet--extensions,
subbundles and tangent spaces, additionally endowed with suitable metrics. For the reader
interested in the general theoretical approach, we refer to \cite{Pasquero2005uni,Pasquero2006}
and the references therein.

\subsection{The geometric environment of the problem}

To study the particular case of two equal billiard balls of radius $r$ and mass $m$ moving on the
billiard plane, we introduce the configuration space--time $\V$, a differentiable manifold fibered
over the time line and parameterized by 11 coordinates. We can choose these coordinates as $(t,
x_1, y_1, \psi_1, \vth_1, \vph_1, x_2, y_2, \psi_2, \vth_2, \vph_2)$, where $t$ is the time
coordinate, $x_i, y_i$ are the coordinates of the center of mass of the $i$-th ball and $\psi_i,
\vth_i, \vph_i$ are the Euler angles giving the orientation of the $i$-th ball. The simultaneous
presence of the balls on the plane implies that not all the configurations of $\V$ are admissible,
since at any instant the coordinates must satisfy the relation $(x_2 - x_1)^2 + (y_2 - y_1)^2 \ge
4r^2$. There is then a natural positional constraint $\S \subset \V$ expressed by the condition
\begin{eqnarray}\label{vincolo posizionale}
\S:= \{(x_2 - x_1)^2 + (y_2 - y_1)^2 - 4r^2 \= 0\}
\end{eqnarray}
that must be satisfied in the moment of the impact (and then called the impact constraint).

The absolute--velocity space--time $\J$ is the first jet--extension of $\V$, parameterized by 21
coordinates: the first 11 are the coordinates running on $\V$, the remaining 10 can be grouped in
a vector ${\bf{p}} \= (1, \dx{1}, \dy{1}, \dotp{1}, \dott{1}, \dotf{1}, \dx{2}, \dy{2}, \dotp{2},
\dott{2}, \dotf{2})$: the presence of the first constant coordinate stresses the necessity of
respecting the affine nature of the velocity space--time.

The space $V(\V)$ describing the possible impulses acting on the system is the vertical fiber
bundle of $\V$. It is a differentiable manifold parameterized by 21 coordinates, whose first 11
are the coordinates running on $\V$ and the remaining 10 can grouped in a vector ${\bf{I}} \= (0,
I_{x_1}, I_{y_1},I_{\psi_1},I_{\vth_1},I_{\vph_1},I_{x_2},
I_{y_2},I_{\psi_2},I_{\vth_2},I_{\vph_2})$: this time the presence of the first constant
coordinate is due to the necessity of respecting the vector nature of the impulse space, and
points out that the fibers of $V(\V)$ are the vector spaces modelling the affine spaces given by
the fibers of $\J$.

The relation between the space of velocities and the space of impulses is cleared by the very
nature of the impact problem: given an input velocity ${\bf{p}}_L$ (the so called
``left--velocity'') of the system and an impulse ${\bf{I}}$, the sum
\begin{eqnarray*}
{\bf{p}}_R = {\bf{p}}_L + {\bf{I}}
\end{eqnarray*}
is required to be an output velocity (``right--velocity'') of the system. A similar situation is
perfectly framed in the relations between affine spaces and modelling vector spaces. However, once
the situation is cleared up of possible confusions, from the computational point of view the
significative parts of velocities ${\bf{p}}$ and impulses ${\bf{I}}$ are given by the 10
components $(\dx{1}, \dy{1}, \dotp{1}, \dott{1}, \dotf{1}, \dx{2}, \dy{2}, \dotp{2}, \dott{2},
\dotf{2})$ and $(I_{x_1}, I_{y_1},I_{\psi_1},I_{\vth_1},I_{\vph_1},I_{x_2},
I_{y_2},I_{\psi_2},I_{\vth_2},I_{\vph_2})$ respectively. These 10--uples will be used without
ulterior specifications in the calculations.

A scalar product $\Phi$ acting on the space of impulses $V(\V)$ is defined by using the positive
definite matrix $g$ that gives the kinetic energy of the system as a quadratic form in the
velocity coordinates $(\dx{1}, \dy{1}, \dotp{1}, \dott{1}, \dotf{1}, \dx{2}, \dy{2}, \dotp{2},
\dott{2}, \dotf{2})$. The elements of $g$ are in general functions of the positional coordinates
$(x_1, y_1, \psi_1, \vth_1, \vph_1, x_2, y_2, \psi_2, \vth_2, \vph_2)$ and take automatically into
account the massive properties of the system. Considering the significative part of the impulse
vectors and denoting with $A$ the inertia momentum of the balls with respect to the center of
mass, we have
\begin{eqnarray}\label{metrica}
g\= \left(
\begin{array}{cccccccccc}
m & 0 & 0 & 0 & 0 & 0 & 0 & 0 & 0 & 0 \\
0 & m & 0 & 0 & 0 & 0 & 0 & 0 & 0 & 0 \\
0 & 0 & A & 0 & A\cos\vth_1 & 0 & 0 & 0 & 0 & 0 \\
0 & 0 & 0 & A & 0 & 0 & 0 & 0 & 0 & 0 \\
0 & 0 & A\cos\vth_1 & 0 & A & 0 & 0 & 0 & 0 & 0 \\
0 & 0 & 0 & 0 & 0 & m & 0 & 0 & 0 & 0 \\
0 & 0 & 0 & 0 & 0 & 0 & m & 0 & 0 & 0 \\
0 & 0 & 0 & 0 & 0 & 0 & 0 & A & 0 & A\cos\vth_2 \\
0 & 0 & 0 & 0 & 0 & 0 & 0 & 0 & A & 0 \\
0 & 0 & 0 & 0 & 0 & 0 & 0 & A\cos\vth_2 & 0 & A
\end{array}
\right) .
\end{eqnarray}
A simple calculation shows that this scalar product is positive definite if and only if
$\sin\vth_1\ne 0$ and $\sin\vth_2 \ne 0$.

\smallskip

Some subspaces of the velocity space--time $\J$ can be naturally introduced. The presence of the
constraint (\ref{vincolo posizionale}) determines the space $\T$ of the velocities that are
tangent to the positional constraint. Straightforward arguments (of Differential or even
Elementary Geometry, since the relative velocities of the centers of mass must be orthogonal to
the radii of the balls at the instant of the impact) show that in our case $\T$ is given by those
velocities obeying the condition
\begin{eqnarray}\label{tangenti vincolo posizionale}
(\dx{2} - \dx{1})(x_2 - x_1) + (\dy{2} - \dy{1})(y_2 - y_1) \= 0 .
\end{eqnarray}
Moreover, the possible presence of rolling conditions
\begin{eqnarray}\label{vincolo cinetico}
\left\{
\begin{array}{l}
\dx{i} - r \dott{i}\sin\psi_i + r \dotf{i} \sin\vth_i\cos\psi_i \= 0 \\
\dy{i} + r \dott{i}\cos\psi_i + r \dotf{i} \sin\vth_i\sin\psi_i \= 0
\end{array}
\right.
\end{eqnarray}
for one or both the balls restricts the space of admissible velocities to a subspace  $\A \subset
\J$. Note that, when the system is subject to (\ref{vincolo cinetico}), the space $\A$ is
restricted before, during and after the impact. However, due to the linearity of (\ref{vincolo
cinetico}), $\A$ still owns the geometric structure of affine space.

\section{Ideality of the unilateral constraint}

The geometric setup synthetically described above allows the introduction of a subspace  $\A \cap
\T \subset \A \subseteq \J $ of the space of admissible velocities $\A$ of the system before and
after the impact given by the admissible velocities of the system when the balls are in contact.
We assume that the contact between the balls is smooth (otherwise an ulterior restriction of the
space of admissible velocities of the system when the balls are in contact must be introduced for
the presence of an ulterior kinetic condition).

Due to the affine structure of these spaces and the existence of the metric structure given by
(\ref{metrica}), for every element ${\bf{p}}$ in $\A$, there exists a closest element
$\P({\bf{p}})$ in $\A \cap \T$. The element $\P({\bf{p}})$ can be thought of as the component of
${\bf{p}}$ tangent to the impact constraint, while the difference
\begin{eqnarray}\label{impulso}
I^{\perp}({\bf{p}}) \= {\bf{p}} - \P({\bf{p}})
\end{eqnarray}
can be thought of as the component of ${\bf{p}}$ orthogonal to the impact constraint.

The most natural behaviour of the impact constraint is then given by a ``reflection''. The impact
with the constraint leaves invariant the tangent component $\P({\bf{p}}_L)$ of the left velocity
${\bf{p}}_L$ and reflects the orthogonal component $I^{\perp}({\bf{p}}_L)$. Then, for every
admissible left velocity ${\bf{p}}_L$, the corresponding right velocity ${\bf{p}}_R$ is given by
\begin{eqnarray}\label{caratterizzazione velocita}
{\bf{p}}_R = {\bf{p}}_L - 2 I^{\perp}({\bf{p}}_L) .
\end{eqnarray}

It can be shown (see \cite{Pasquero2005uni, Pasquero2006}) that the right velocity ${\bf{p}}_R$
given by (\ref{caratterizzazione velocita}) belongs to $\A$ if ${\bf{p}}_L\in\A$. Moreover, it
preserves the kinetic energy in the impact (for every observer for which the constraint is at
rest, that are the only ones for which the conservation of kinetic energy has a clear meaning
\cite{Pasquero2005Carnot}). This property motivates the ideality of the chosen reaction.

\section{The algorithm}

The algorithm to determine the right velocity ${\bf{p}}_R $ starting from a left velocity
${\bf{p}}_L$ can be easily described by the following procedure:

\begin{itemize}
\item[1)] we start from an {\it a priori} knowledge of the space of admissible velocities $\A$
(coinciding with the whole $\J$ when no rolling constraints act on  the balls), its subspace $\A
\cap \T$ (determined by the impulse kinetic constraint \ref{tangenti vincolo posizionale}) and the
scalar product \ref{metrica};

\item[2)] we have as input an assigned left--velocity ${\bf{p}}_L \in \A$;

\item[3)] we determine the tangent component $\P({\bf{p}}_L) \in \A \cap \T$ and the orthogonal
component $I^{\perp}({\bf{p}}_L)$;

\item[4)] we determine the right--velocity ${\bf{p}}_R \in \A$ using the relation ${\bf{p}}_R =
{\bf{p}}_L - 2 I^{\perp}({\bf{p}}_L)$.

\end{itemize}

Clearly, from the operative point of view, the core of the algorithm is point 3): it consists in
the solution of the Least Square Problem determining the projection $\P({\bf{p}}_L)$ of the
assigned left velocity ${\bf{p}}_L \in \A$ onto the space $\A\cap \T$, together with the condition
(\ref{tangenti vincolo posizionale}) in the smooth case, together with the conditions
(\ref{tangenti vincolo posizionale}) and (\ref{vincolo cinetico}) for both $i=1,2$ in the rolling
case, and together with the conditions (\ref{tangenti vincolo posizionale}) and (\ref{vincolo
cinetico}) for only one $i \in \{1,2\}$ for the mixed case.

\smallskip

Starting from the $10$--uple ${\bf{p}}_L \simeq (\dx{1}^L, \dy{1}^L, \dotp{1}^L, \dott{1}^L,
\dotf{1}^L, \dx{2}^L, \dy{2}^L, \dotp{2}^L, \dott{2}^L, \dotf{2}^L)$, we introduce the function of
the $10$ unknown $(\dx{1}, \dy{1}, \dotp{1}, \dott{1}, \dotf{1}, \dx{2}, \dy{2}, \dotp{2},
\dott{2}, \dotf{2})$
\begin{eqnarray}
\begin{array}{lcl}
\| {\bf{p}} -  {\bf{p}}_L \| ^2 &\=& m \left( \dx{1} - \dx{1}^L
\right)^2 \, + \, m \left( \dy{1} - \dy{1}^L \right)^2  \\
 &&\, + \, A \left(  \dotp{1} - \dotp{1}^L \right)^2 \, + \, A \left( \dott{1} - \dott{1}^L
\right)^2
 + \, A \left( \dotf{1} - \dotf{1}^L \right)^2  \, \\
&& \, + \, 2\, A \cos\psi_1 \left( \dotp{1} -
\dotp{1}^L  \right)\left( \dotf{1} - \dotf{1}^L \right)  \\
&& \, + \, m \left( \dx{2} - \dx{2}^L \right)^2 \, + \, m \left(
\dy{2} - \dy{2}^L \right)^2  \\
&& \, + \, A \left(  \dotp{2} - \dotp{2}^L \right)^2 \, + \, A
\left( \dott{2} - \dott{2}^L \right)^2 \, + \, A \left( \dotf{2} -
\dotf{2}^L \right)^2  \\
&& \, \, + \, 2\, A \cos\psi_2 \left( \dotp{2} - \dotp{2}^L
\right)\left( \dotf{2} - \dotf{2}^L \right).
\end{array}
\end{eqnarray}
Using the Lagrange Multipliers Method, in the smooth case we look
(neglecting the parameters introduced by the LMM) for a $10$--uple
minimizing the function
\begin{eqnarray}
\begin{array}{lcl} {\cal L}_{smooth} &\=& \| {\bf{p}} -  {\bf{p}}_L \| ^2 \, + \, \lambda \,
\left[ (\dx{2} - \dx{1})(x_2 - x_1) + (\dy{2} - \dy{1})(y_2 - y_1) \right] .
\end{array}
\end{eqnarray}
In the rolling case we look (once again neglecting the parameters
introduced by the LMM) for a $10$--uple minimizing the function
\begin{eqnarray}
\begin{array}{lclll} {\cal L}_{roll} &\=& \| {\bf{p}} -  {\bf{p}}_L \| ^2 \, &+& \, \lambda \,
\left[ (\dx{2} - \dx{1})(x_2 - x_1) + (\dy{2} - \dy{1})(y_2 - y_1) \right] \\ && \, &+& \,
\mu_{11} \, \left( \dx{1}
- r \dott{1}\sin\psi_1 + r \dotf{1} \sin\vth_1\cos\psi_1 \right) \\
&& \, &+& \, \mu_{12} \, \left( \dy{1} + r \dott{1}\cos\psi_1 +
r \dotf{1} \sin\vth_1\sin\psi_1 \right) \\
&& \, &+& \, \mu_{21} \, \left( \dx{2} - r \dott{2}\sin\psi_2 + r
\dotf{2} \sin\vth_2\cos\psi_2 \right) \\ && \, &+& \, \mu_{22} \,
\left( \dy{2} + r \dott{2}\cos\psi_2 + r \dotf{2}
\sin\vth_2\sin\psi_2 \right) .
\end{array}
\end{eqnarray}
In the mixed case (supposing the ball nr. $2$ rolling without
sliding) we look for a $10$--uple minimizing the function
\begin{eqnarray}
\begin{array}{lclll}
{\cal L}_{mixed} &\=& \| {\bf{p}} -  {\bf{p}}_L \| ^2 \, &+& \, \lambda \, \left[ (\dx{2} -
\dx{1})(x_2 - x_1) + (\dy{2} - \dy{1})(y_2 - y_1) \right] \\ && \, &+& \, \mu_{21} \, \left(
\dx{2}
- r \dott{2}\sin\psi_2 + r \dotf{2} \sin\vth_2\cos\psi_2 \right) \\
&& \, &+& \, \mu_{22} \, \left( \dy{2} + r \dott{2}\cos\psi_2 + r
\dotf{2} \sin\vth_2\sin\psi_2 \right) .
\end{array}
\end{eqnarray}

In all cases the minimizing $10$--uple gives the required projection $\P({\bf{p}}_L)$, and then
the main object for the analysis of the impact.

\section{The possible ideal impacts}

In  this section we present the results of the computation in the
three cases described above. To perform this computation, we used
CoCoA 4.7, a freeware software for polynomial symbolic computation
(\cite{CoCoA}). For each situation, we present some general (in the
sense of ``depending by the whole set of parameters'') but
significative results, and some results obtained for particular
impacts (in the sense of ``having chosen some parameters in a simple
but significative way'').

\subsection{Case I. The smooth situation}

It can be easily proved that, when the contact between the balls and the plane of the billiard is
smooth, the impact does not affect the behaviour of the ``angle'' variables $(\dotp{1}, \dott{1},
\dotf{1}, \dotp{2}, \dott{2}, \dotf{2})$ and the unique variations regard the ``linear'' variables
$(\dx{1}, \dy{1}, \dx{2}, \dy{2})$ of the velocities of the centers of mass of the balls. In
particular, we have the following results:
\begin{eqnarray}\label{velocita destre lisce}
\begin{array}{lcl}
\dx{1}^R &\=& \dfrac{\dx{1}^L(y_2 - y_1)^{2} + \dx{2}^L(x_2 -
x_1)^{2}  + (\dy{2}^L - \dy{1}^L)(x_2 - x_1)(y_2 - y_1)
}{4r^{2}} \\ \\
\dy{1}^R &\=&  \dfrac{\dy{1}^L(x_2 - x_1)^{2} +
\dy{2}^L(y_2 - y_1)^{2} + (\dx{2}^L - \dx{1}^L)(x_2 - x_1)(y_2 - y_1) }{4r^{2}} \\ \\
\dx{2}^R &\=& \dfrac{\dx{1}^L(x_2 - x_1)^{2} +
\dx{2}^L(y_2 - y_1)^{2} - (\dy{2}^L - \dy{1}^L)(x_2 - x_1)(y_2 - y_1) }{4r^{2}} \\ \\
\dy{2}^R &\=& \dfrac{\dy{1}^L(y_2 - y_1)^{2} + \dy{2}^L(x_2 -
x_1)^{2} - (\dx{2}^L - \dx{1}^L ) (x_2 - x_1)(y_2 - y_1)}{4r^{2}}.
\end{array}
\end{eqnarray}
Note that, while the exit velocity depends on the relative position
of the centers of mass of the balls at the impact, it does not
depend on the orientations of the balls at the impact.

Let us denote with $\underline{\bf{V}}_i^L, \underline{\bf{V}}_i^R, i=1,2$ the ``usual'' left and
right velocities of the centers of mass of the balls (so that, for example, the component of
$\underline{\bf{V}}_1^L$ along the $x$ direction is $\dx{1}^L$) and with $\underline{\bf u}
\cdot\underline{\bf w}$ the ``usual'' scalar product. Then the results (\ref{velocita destre
lisce}) satisfy the relation
\begin{eqnarray}\label{prodotto scalare destro liscio}
\underline{\bf{V}}_1^L \cdot \underline{\bf{V}}_2^L \= \underline{\bf{V}}_1^R \cdot
\underline{\bf{V}}_2^R .
\end{eqnarray}
In particular (\ref{prodotto scalare destro liscio}) implies that,
when the impact happens with one of the two balls at rest,
independently of the impact angle, the exit velocities of the two
centers of mass are orthogonal.

The computation shows also that the reactive impulse $-2 I^{\perp}({\bf p}_L)$ has (null
components relative to the ``angle'' variables and) components relative to the ``linear''
variables
\begin{eqnarray}\label{impulsi lisci}
\begin{array}{l}
I^{\perp}_{x_1} = - I^{\perp}_{x_2} = - \dfrac{ ( \dx{2}^L -
\dx{1}^L )  (x_2 - x_1)^{2}  + ( \dy{2}^L - \dy{1}^L
  ) (x_2 - x_1)(y_2 - y_1)}{2r^{2}} \propto (x_2 - x_1)
\\ \\
I^{\perp}_{y_1} = - I^{\perp}_{y_2} = -  \dfrac{ ( \dx{2}^L -
\dx{1}^L     )(x_2 - x_1)(y_2 - y_1) + ( \dy{2}^L  - \dy{1}^L) (y_2
- y_1)^{2}}{2r^{2}} \propto (y_2 - y_1) .
\end{array}
\end{eqnarray}
Therefore, each reactive impulse relative to a single ball turns out to be parallel to the
direction determined by the centers of mass of the balls. This property, together with the
smoothness of the contact between balls and plane), motivate also why the impact does not affect
the ``spin'' of the balls (see \cite{SalaSanc1990}).

Moreover, simple tests show the reasonableness of (\ref{velocita destre lisce}). When
$\underline{\bf{V}}_1^L \= 0$, and the impact is direct central, then $\underline{\bf{V}}_2^R\=0,
\underline{\bf V}_1^R \= \underline{\bf{V}}_2^L$. More generally, when $\underline{\bf{V}}_1^L \=
0$, then $\underline{\bf{V}}_1^R$ turns out to be parallel to the direction determined by the
centers of mass at the impact.

\subsection{Case II. The pure rolling situation}

It can be easily foreseen that, differently from the smooth case,
when both the balls roll without sliding on the billiard plane, in
general all the $10$ component representing the velocities of the
balls change with the impact. The computation support this forecast.
In particular, the ''linear'' velocity variables have the same
behaviour in the smooth and the rolling cases, so that
(\ref{velocita destre lisce}) still hold, together with
(\ref{prodotto scalare destro liscio}, \ref{impulsi lisci}).

Unfortunately, although explicitly computable,  the changes of the ``angle'' velocity variables
have complicated expressions, without easy interpretation. For example, we have,
\begin{eqnarray*}
\begin{array}{lcl}
\dott{1}^R &\=&  \dfrac{1}{4r^{3}}\Big\{ ( \dx{2}^L\sin\psi_1 -
\dy{1}^L\cos\psi_1 )  (x_2 - x_1)^{2}
\\ \\ && +  \Big[ r \dott{1}^L - ( \dy{2}^L - \dy{1}^L ) \Big] \cos\psi_1(y_2 - y_1)^{2}
\\ \\ && + \Big[  (\dy{2}^L-\dy{1}^L)\sin\psi_1  - (\dx{2}^L - \dx{1}^L) \cos\psi_1    \Big](x_2 - x_1)(y_2 - y_1)
 \Big\} .
\end{array}
\end{eqnarray*}
It is then convenient to restrict our analysis to particular situations, such as when the left
velocity of one ball is fixed, and possibly null, and fixing the angle between
$\underline{\bf{V}}_2^L $ and the direction determined by the centers of mass.

For example, for an impact with the ball nr. 1 at rest, $\underline{\bf{V}}_2^L$ parallel to the
$x$-direction and angle between $\underline{\bf{V}}_2^L $ and the direction determined by the
centers of mass of $\frac{\pi}4$, we have
\begin{eqnarray*}
\qquad \qquad
\begin{array}{l}
\dx{1}^R \= \dfrac12 \dx{2}^L \, ; \\ \\   \dy{1}^R \= \dfrac12 \dx{2}^L \, ;
\\ \\
\dotp{1}^R \= \dfrac{( \sin\psi_1 + \cos\psi_1)\cos\vth_1}{2\sin\vth_1} \,  \dfrac{\dx{2}^L}{r} \,
; \\ \\
\dott{1}^R \= \dfrac{( \sin\psi_1 - \cos\psi_1)}{2} \, \dfrac{\dx{2}^L}{r} \, ;
\\ \\
\dotf{1}^R \= - \dfrac{( \sin\psi_1 + \cos\psi_1)}{2 \sin\vth_1} \, \dfrac{\dx{2}^L}{r} \, ;
\\ \\
\dx{2}^R \= \dfrac12 \dx{2}^L  \, ; \\ \\  \dy{2}^R \= -\dfrac12 \dx{2}^L  \, ;
\\ \\
\dotp{2}^R \= \dotp{2}^L - \dfrac{\cos\vth_2}{\sin\vth_2} \, \dfrac{ r\dott{2}^L +
\dx{2}^L\cos\psi_2
 }{2r} \, ;
\\ \\
\dott{2}^R \= \dfrac{ r\dott{2}^L + \dx{2}^L\cos\psi_2 }{2r}  \, ;
\\ \\
\dotf{2}^R \= \dfrac{1}{\sin\vth_2} \, \dfrac{r\dott{2}^L - \dx{2}^L\cos\psi_2}{2r} .
\end{array}
\end{eqnarray*}

Another simple example is given by the direct central impact with the ball nr. 1 at rest and
$\underline{\bf{V}}_2^L$ parallel to the $x$-direction. We have
\begin{eqnarray*} \qquad \qquad
\begin{array}{l}
\dx{1}^ R \= \dx{2}^L ; \, \\ \\  \dy{1}^R \= 0
\\ \\
\dotp{1}^R \= \dfrac{\cos\psi_1\cos\vth_1}{\sin\vth_1} \, \dfrac{
\dx{2}^L}{r}
 ; \end{array}
\end{eqnarray*}
\begin{eqnarray*} \qquad \qquad
\begin{array}{l}
\dott{1}^R \=  \sin\psi_1 \, \dfrac{\dx{2}^L}{r} ; \\ \\  \dotf{1}^R \=  -
\dfrac{\cos\psi_1}{\sin\vth_1} \, \dfrac{\dx{2}^L}{r}
\\ \\
\dx{2}^ R \= 0 ; \\ \\  \dy{2}^ R \=  0
 \\ \\
\dotp{2}^R \= \dotp{2}^L - \dfrac{\cos\psi_2\cos\vth_2}{\sin\vth_2} \, \dfrac{  \dx{2}^L}{r} ;
\\ \\  \dott{2}^R \=  0 ; \\ \\  \dotf{2}^R \=  0 .
\end{array}
\end{eqnarray*}

\subsection{Case III. The mixed situation}

The mixed situation, when one of the balls rolls without sliding and the other slides on the
billiard plane seems a very artful situation. However, taking into account the results presented
in \cite{Torre1994} about the behaviour of a billiard ball moving with friction on the billiard
plane, the case of a rolling ball impacting with another ball at rest so that the second ball
starts to move sliding on the billiard plane is the ideal case that better models a ``real world''
situation. Once again, some simple particular impacts give  clearer outlooks than general results.
For example, let us consider once again an impact with the ball nr. 1 at rest (on a smooth part of
the billiard plane), the ball nr. 2 rolling on the billiard plane with $\underline{\bf V}_2^L$
parallel to the $x$-direction and angle between $\underline{\bf{V}}_2^L $ and the direction
determined by the centers of mass of $\frac{\pi}4$. If we set $A \= a m r^2$, we obtain
\begin{eqnarray*}
\qquad \qquad
\begin{array}{l}
\dx{1}^R \= \dfrac{a + 1}{a + 2} \, \dx{2}^L \, ; \\ \\  \dy{1}^R \= \dfrac{a + 1}{a + 2} \,
\dx{2}^L \, ;
\\ \\
\dotp{1}^R \=   0 \, ; \\ \\   \dott{1}^R \=  0 \, ; \\ \\  \dotf{1}^R \= 0
\\ \\
\dx{2}^R \=  \dfrac{a + 1}{a + 2} \, \dx{2}^L \, ; \\ \\ \dy{2}^R \=  - \dfrac{1}{a + 2} \,
\dx{2}^L
\\ \\
\dotp{2}^R \=  \dotp{2}^L - \dfrac{1}{a + 2} \, \dfrac{\cos\vth_2}{\sin\vth_2} \,  \left(
\dott{2}^L + \dfrac{\dx{2}^L}{r}\cos\psi_2 \right) \, ;
\\ \\
\dott{2}^R \= \dfrac{1 }{a + 2}\, \left( (a + 1) \, \dott{2}^L + \, \dfrac{ \dx{2}^L}{r} \,
\cos\psi_2  \right) \, ;
\end{array}
\end{eqnarray*}
\begin{eqnarray*} \qquad \qquad
\begin{array}{l}
\dotf{2}^R \= \dfrac{1}{a+2} \, \dfrac{1}{\sin\vth_2} \, \left( \dott{2}^L - (a+1)  \, \dfrac{
\dx{2}^L}{r} \, \cos\psi_2\right) .
\end{array}
\end{eqnarray*}
However, in this case the most interesting result is given by the scalar product
$\underline{\bf{V}}_1^R \cdot \underline{\bf{V}}_2^R$ of the right ``linear'' velocities: a
straightforward calculation shows that
\begin{eqnarray*}
\underline{\bf{V}}_1^R \cdot \underline{\bf{V}}_2^R \= \dfrac{a (a+1) }{(a+2)^2} \, (\dx{2}^L)^{2}
\ne 0 .
\end{eqnarray*}
Therefore, in this situation, the exit velocities of the balls are
not orthogonal. Since this is, for every billiard player, a very
well known ``real'' fact, the result gives an ulterior proof of the
good agreement of the result of the mixed model to the real
situation.

\section{Conclusions}

We used a symbolic computation software to obtain detailed results in the study of the behaviour
of two equal billiard balls colliding in three ideal situations. The use of computation software
was almost essential in order to obtain effective results by the known theoretical procedure,
since the high number of degrees of freedom of the system involves calculations that could hardly
be approached by hand techniques.

Although the computation were performed in very general conditions, often the results obtained can
be interpreted only for particular impacts, such as when one of the ball is at rest. Therefore,
only a few of the numerous interesting impacts where analyzed in details.

The results in the smooth and the pure rolling situations respect the forecasts (easier in the
smooth case, more complicated in the other), especially about the angle formed by the directions
of the balls after the impact, and about the spins of the balls after the impact. The analysis of
the mixed situation, suggested by similar analyses of analogous problems that can be found in
literature, although the apparent artfulness of the case, give results that are in good agreement
to well known real behaviours of the billiard balls.



\end{document}